\documentclass[journal]{IEEEtran}

\usepackage{ amsmath, bm, hyperref, authblk, nomencl, lipsum, chngcntr, blindtext}
\usepackage{tikz}
\usepackage[utf8]{inputenc}
\usepackage{pbox}
\usepackage{cite}
\usepackage[english]{babel}

\usepackage{booktabs}
\usepackage{placeins}
\usepackage{forest}
\usetikzlibrary{arrows,decorations.pathmorphing,backgrounds,fit,positioning,shapes.symbols,chains, shapes,decorations}
\usepackage{csquotes}
\usepackage{comment}
\usepackage{multirow}
\usepackage{colortbl}
\usepackage{color}
\usepackage{amsfonts}
\usepackage{hyperref}
\usetikzlibrary{arrows.meta}
\usetikzlibrary{shapes,arrows}
\usetikzlibrary{automata, positioning}
\usepackage[makeroom]{cancel} 

\usetikzlibrary{arrows,decorations.pathmorphing,backgrounds,fit,positioning,shapes.symbols,chains, shapes,decorations}

\title{Optimal Load Ensemble Control in Chance-Constrained Optimal Power Flow}
\author{Ali Hassan, Robert Mieth, Michael Chertkov, Deepjyoti Deka\thanks{The authors acknowledge support from NSF \# CMMI-1825212 and DOE through the Grid Modernization Lab Consortium, and the Center for Non Linear Studies (CNLS) at Los Alamos.}, Yury Dvorkin

} 	
\date{} 

\begin{document}			
\clearpage
\thispagestyle{empty}
\maketitle

\begin{abstract}
Distribution system operators (DSOs) world-wide foresee a rapid roll-out of distributed energy resources. From the system perspective, their reliable and cost effective integration requires accounting for their physical properties in operating tools used by the DSO. This paper describes an decomposable approach to leverage the dispatch flexibility of thermostatically controlled loads (TCLs) for operating distribution systems with a high penetration level of photovoltaic resources. Each TCL ensemble is modeled using the Markov Decision Process (MDP). The MDP model is then integrated with a chance constrained optimal power flow that accounts for the uncertainty of PV resources. Since the integrated optimization model cannot be solved efficiently by existing dynamic programming methods or off-the-shelf solvers, this paper proposes an iterative Spatio-Temporal Dual Decomposition algorithm (ST-D2). We demonstrate the merits of the proposed integrated optimization and ST-D2 algorithm on the IEEE 33-bus test system.

\end{abstract}

\section{Introduction} \label{Sec:introduction}

Distributed energy resources (DERs) are viewed as a techo-economically viable alternative to conventional generation resources and, in some cases, have been shown to provide cost-competitive system support services, including peak shaving, ancillary services, emergency and disaster response, and investment deferral \cite{61030}. E.g., the State of New York estimates a total technical potential of roof-top photovoltaic (PV) resources (roughly, 80\% of DERs) at 2,615 MW of the cumulative peak capacity and 8,223 GWh production by 2030 \cite{nyiso_report}. Under such PV penetration levels, distribution system operators (DSO) are likely to exhaust existing means for compensating PV intermittency, as well as for distribution power flow and voltage control. In turn, lack of such means may limit the ability to further integrate DER resources. One way to deal with such challenges and overcome existing barriers for DER integration is to fully realize the potential of behind-the-meter demand response capabilities \cite{nyiso_report}. This paper proposes an approach to leverage the flexibility of behind-the-meter thermostatically controlled loads (TCLs) for operating PV-dominant distribution systems.

Previously, demand response capabilities have been considered at the system-level for centralized, hierarchical, and distributed control architectures \cite{Callaway_control}. The common element of these architectures is their reliance on load aggregators as mediators between the DSO and behind-the-meter DERs that are not observable by the DSO. Each aggregator can continuously refine knowledge of each individual resource, e.g. via machine learning \cite{7401112}, and use this knowledge to accurately quantify their dispatch capabilities. In turn, this flexibility whether in the form of load curtailment or load increase can be offered by the aggregator to the DSO in exchange for a certain compensation. The primary difficulty associated with such schemes is two-fold. First, the aggregator needs to model an ensemble of individual TCLs that may vary in a wide range. Second, these resources are driven by the activity of their customers and therefore the aggregator needs to accurately predict their behavior. In case of TCL ensembles, these two difficulties can simultaneously be overcome by modeling each ensemble as a Markov Decision Process (MDP). In \cite{TCL_Angeli, TCL_Mathieu, Chertkov_MDP, TCL_Meyn,TCL_CALLAWAY}, each TCL ensemble is modeled as a discrete-time, discrete-space MDP that is well suited for capturing stochastic dynamics of individual TCLs and is computationally scalable to accommodate hundreds of TCLs in each ensemble. The models \cite{Chertkov_MDP,TCL_Meyn,TCL_CALLAWAY} exploit naive economic dispatch frameworks that co-optimize the flexibility of TCL ensembles and distribution system operations. The common caveat of \cite{Chertkov_MDP,TCL_Meyn,TCL_CALLAWAY} is that network constraints are neglected and, as a result, these models do not ensure compliance with power flow and voltage limits. Furthermore, \cite{Chertkov_MDP,TCL_Meyn,TCL_CALLAWAY} do not account for the uncertainty of PV injections and treat these resources in a deterministic manner. The former caveat is addressed in \cite{TCL_networks_Misha}, where TCL ensembles are operated by the aggregator and network constraints are included using the \textit{LinDistFlow }power flow model \cite{Lindistflow}. The resulting problem in \cite{TCL_networks_Misha} is solved using an iterative solution technique. Similarly to \cite{Chertkov_MDP,TCL_Meyn,TCL_CALLAWAY}, the model in \cite{TCL_networks_Misha} treats PV injections deterministically and therefore is vulnerable to the effects of their intermittency.

Recent efforts to account for the intermittency of PV resources in decision-support tools for DSOs include the use of stochastic programming, especially chance constraints \cite{CCopf_Emiliono, Robert_cdc_2018, 6532318, Hassan_Chance_2017}. Chance constraints naturally fit distribution system operations as they impose a desired tolerance (probability level) to violations of technical constraints (e.g. power flow and voltage limits) so that DSOs can adjust their tolerance based on their reliability preferences and standards. Additionally, the use of chance constraints is motivated as follows. First, as in \cite{CCopf_Bienstock}, they can be reformulated as second-order conic (SOC) constraints that are computationally tractable. Such reformulations exist for multiple probability distributions that are shown to accurately represent the uncertainty of PV resources \cite{6652224}. Second, chance constraints make it possible to trade-off solution cost and robustness by adjusting the desired tolerance to constraint violations. Finally, chance constraints have a well-established connection to data-driven optimization methods, \cite{Jiang2016}, that can be leveraged to overcome limitations of assuming a particular probability distribution. Dall'Anese \textit{et al.} \cite{CCopf_Emiliono} present a chance-constrained (CC) optimal power flow (CC-OPF) model with AC power flow constraints based on the \textit{LinDistFlow} power flow model. This work is extended in \cite{Hassan_Chance_2017} by introducing new power-flow-based control policies for PV~resources that enhance their ability to participate in voltage regulation and power loss minimization. Reference \cite{Robert_cdc_2018} extends the chance constraints derived in \cite{CCopf_Emiliono, Hassan_Chance_2017} under the Gaussian assumption into a data-robust form. \textcolor{black}{Reference \cite{li2017chance} optimizes the TCL dispatch in a centralized manner using the chance-constrained framework. However, solving this optimization from the perspective of the centralized controller increases computational complexity of the problem  and, therefore, it may not scale well due to a significant communication overhead.}

\textcolor{black}{With the exception of our previous work in \cite{TCL_networks_Misha}, the MDP based model for TCLs and CC-OPF optimization are always performed separately.} Since TCL ensembles and PV resources are best modeled by the MDP and chance constrained frameworks respectively, this paper seeks to bridge the gap between the MDP approaches to model TCL ensembles from the aggregator perspective, \cite{Chertkov_MDP,TCL_Meyn,TCL_CALLAWAY}, and the CC-OPF literature that operates the distribution system from the centralized DSO perspective, \cite{CCopf_Emiliono, Robert_cdc_2018, 6532318, Hassan_Chance_2017}. \textcolor{black}{Based on \cite{TCL_networks_Misha}, we propose a decomposition-based algorithm that divides the optimization tasks between the DSO and the TCL ensembles, while minimizing communication needs among them. This paper makes the following contributions:}

\begin{enumerate}
  \item It formulates an integrated optimization problem that includes both the MDP optimization of TCL ensembles and the CC-OPF optimization of the distribution system. Relative to the previous work in \cite{Chertkov_MDP,TCL_Meyn,TCL_CALLAWAY, CCopf_Emiliono, Robert_cdc_2018, 6532318, Hassan_Chance_2017}, the integrated model \textcolor{black}{not only accounts for the TCL dispatch, \textcolor{black}{but also} ensures compliance with distribution system limits and internalizes the PV uncertainty via chance constraints. Furthermore, the CC-OPF optimization is extended to account for the expected value of the quadratic real power losses in the objective function. }
  \item \textcolor{black}{To efficiently solve the proposed integrated model, we develop a Spatio-Temporal Dual Decomposition (\mbox{ST-D2}) algorithm, which is based on the traditional dual decomposition \cite{ADMM}. This novel application of the dual decomposition makes it possible to co-optimize the MDP and CC-OPF subproblems  iteratively using dynamic programming and SOC programming methods, respectively.}
\end{enumerate}
\textcolor{black}{The proposed integrated model and ST-D2 algorithm are tested on the IEEE 33-bus test system presented in \cite{33_bus} to demonstrate the efficiency of our model. To assess the performance of the proposed algorithm and demonstrate its scalability, additional simulations are performed on the 37-, 123- and 8500-bus IEEE systems \cite{schneider2017analytic}.}

The rest of the paper is organized as follows. Section \ref{Sec:formulation} presents an MDP model for optimizing the dispatch of TCL ensembles operated by the aggregator and then integrates this model with the distribution CC-OPF model. \textcolor{black}{Section \ref{Sec:ST} describes the proposed algorithm to solve the integrated model described in Section \ref{Sec:formulation}.} Section \ref{Sec:Case_study} presents the case study to validate the the proposed model and algorithm. Section \ref{Sec:Conclusion} concludes the paper.

\section{Mathematical Formulation} \label{Sec:formulation}
\subsection{Preliminaries}
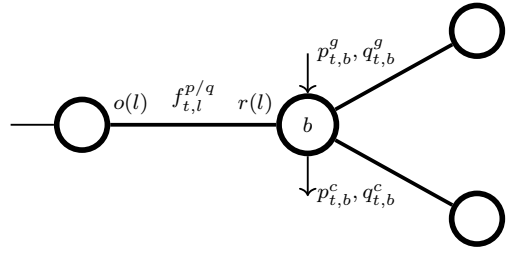
\begin{figure}[!t]
\centering
  \begin{tikzpicture}[auto, node distance=0.5cm,auto, font=\footnotesize]
  \tikzstyle{block} = [circle, draw, fill=white!15, text width=0.4cm, text centered, minimum height=0.4cm]
  \tikzstyle{block_rect} = [rectangle, draw, dotted, fill=white!10, text width=1.5cm, text centered, minimum height=1.5cm]
  
  \node [block, line width=0.75mm] (b1) {} ;
  \node [block, right of=b1, xshift=2.50cm, line width=0.75mm ] (b2) {$b$} ;
  \node [block, right of=b2, xshift=1.75cm, yshift=1.25cm, line width=0.75mm ] (b2a) {} ;
  \node [block, right of=b2, xshift=1.75cm, yshift=-1.25cm, line width=0.75mm ] (b2b) {} ;

	\coordinate[below of=b1, yshift=-0.45cm] (c1);
	\coordinate[left of=b1, xshift=-0.45cm] (c2);
	
	\coordinate[below of=b2, yshift=-0.45cm] (c3);
	\coordinate[right of=b2, xshift=-0.45cm] (c4);
	\coordinate[above of =b2, yshift = 0.45cm] (b9);
	
  \draw [-,thick, line width=0.5mm] (b1) -- (b2) node [very near start] {$o(l)$} node [near start] {$\qquad\qquad f_{t,l}^{p/q}$} node [very near end] {$r(l)$};
  \draw [-,thick, line width=0.25mm] (c2) -- (b1) node [near start] {};
  \draw [->,thick, line width=0.25mm] (b2) -- (c3) node [at end] {$p_{t,b}^{c},q_{t,b}^{c}$};
   \draw [->,thick, line width=0.25mm] (b9) -- (b2) node [at start] {$p_{t,b}^{g},q_{t,b}^{g}$};
  \draw [-,thick, line width=0.5mm] (b2) -- (b2a) node [midway,sloped,anchor=south] {} node [very near start,sloped,anchor=south] {};
  \draw [-,thick, line width=0.5mm] (b2) -- (b2b) node [midway,sloped,anchor=south] {} node [very near start,sloped,anchor=south] {};
\end{tikzpicture}\textbf{}
\vspace{3pt}
\caption{A radial distribution system with notations.}
\label{fig:notations}
\end{figure}

We consider a radial distribution system represented by graph $\Gamma$ = $(\mathcal{E},\mathcal{N})$, where $\mathcal{E}$ and $\mathcal{N}$ are the sets of lines (edges) and buses (nodes), see Fig.~\ref{fig:notations}. The set of operating time intervals is represented by $\mathcal{T}$, indexed by $t$. The set of buses where controllable generation resources are located is denoted as $\mathcal{N}^{\mathcal{G}} \subseteq \mathcal{N}$, the set of buses where PV resources are located is denoted as $\mathcal{N}^{\mathcal{P}\mathcal{V}} \subseteq \mathcal{N}$ and the set of buses where TCL ensembles are located is denoted as $\mathcal{N}^{\mathcal{T}} \subseteq \mathcal{N}$. Each node can be characterized by its active and reactive load ($p_{t,b}^c$ and $q_{t,b}^c$, $\forall t$, $b \in \mathcal{N}$), active and reactive power output of controllable generation resources ($p_{t,b}^g$ and $q_{t,b}^g$, $\forall t$, $b \in \mathcal{N}^{\mathcal{G}}$), active and reactive PV generation ($p_{t,b}^{\mathcal{PV}}$ and $q_{t,b}^{\mathcal{PV}}$, $\forall t$, $b \in \mathcal{N}^{\mathcal{PV}}$), active and reactive injections ($p_{t,b}$ and $q_{t,b}$, $\forall t$, $b \in \mathcal{N}^{\mathcal{T}}$) and voltage magnitude $v_{t,b} \in$[$\overline{V}_b,\underline{V}_b$], $\forall t$, $b\in\mathcal{N}$, where $\overline{V}_b$ and $\underline{V}_b$ are the upper and lower nodal voltage limits respectively. The square of the nodal voltage is denoted as $u_{t,b}$ = $v_{t,b}^2$, $\forall t$, $b\in \mathcal{N}$, with limits as $\overline{U}_b$ = $(\overline{V}_b)^2$ and $\underline{U}_b$ = $(\underline{V}_b)^2$. Each line is characterized by its active and reactive power flows ($f_{t,l}^p$ and $f_{t,l}^q$, $\forall t$, $l\in \mathcal{E}$), its resistance and reactance ($R_{l}$ and $X_{l}$, $l\in \mathcal{E}$). The origin and receiving buses for each line are indexed as $o(l)$ and $r(l)$. The bold font will denote the uncertain quantities.

\textcolor{black}{The main notations are defined as follows:} 
\subsection*{\textcolor{black}{Variables}}
\begin{IEEEdescription}[\IEEEusemathlabelsep\IEEEsetlabelwidth{ve long}]
\item[\textcolor{black}{$f_{t,l}^{p/q}$}]{\textcolor{black}{Active/reactive power flows in line $l$ during time interval $t$}}
\item[\textcolor{black}{${\mathcal{P}}_{t,b}^{\alpha\beta}$}]{\textcolor{black}{Transition probability from state $\beta$ to state $\alpha$ for the TCL ensemble at bus $b$ during time interval $t$}}
\item[\textcolor{black}{$p_{t,b}/q_{t,b}$}]{\textcolor{black}{Active/reactive power injections at bus $b$ during time interval $t$}}
\item[\textcolor{black}{$p_{t,b}^{g}/q_{t,b}^{g}$}]{\textcolor{black}{Active/reactive power output of controllable generation resources at bus $b$ during time interval $t$}}
\item[\textcolor{black}{$p_{t,b}^{\mathcal{P}\mathcal{V}}/q_{t,b}^{\mathcal{P}\mathcal{V}}$}]{\textcolor{black}{Active/reactive PV generation at bus $b$ during time interval $t$}}
\item[\textcolor{black}{$p_{b}^{\alpha}/q_{b}^{\alpha}$}]{\textcolor{black}{Rated active/reactive power consumption at state $\alpha$ for the TCL ensemble at bus $b$}}
\item[\textcolor{black}{$u_{t,b}$}]{\textcolor{black}{Voltage magnitude squared at bus $b$ during time interval $t$}}
\item[\textcolor{black}{$v_{t,b}$}]{\textcolor{black}{Voltage magnitude at bus $b$ during time interval $t$}}
\item[\textcolor{black}{$\alpha_{t,b}$}]{\textcolor{black}{Participation factor of the controllable generation resource at bus $b$ during time interval $t$}}
\item[\textcolor{black}{$\gamma_{t,b}^{\alpha\beta}$}]{\textcolor{black}{Cost penalty factor on the transition from state $\alpha$ to state $\beta$ for the TCL ensemble at bus $b$ during time interval $t$}}
\item[\textcolor{black}{$\lambda_{t,b}^{p/q}$}]{\textcolor{black}{Lagrange multipliers}}
\item[\textcolor{black}{$\rho_{t,b}^{\alpha/\beta}$}]{\textcolor{black}{Probability of the TCL ensemble being in state $\alpha/\beta$ located at bus $b$ during time interval $t$}}
\end{IEEEdescription}

\subsection*{\textcolor{black}{Parameters}}
\begin{IEEEdescription}[\IEEEusemathlabelsep\IEEEsetlabelwidth{ve long}]
\item[\textcolor{black}{$\overline{G}_{b}^p/\underline{G}_{b}^p$}]{\textcolor{black}{Maximum/minimum active power output of controllable generation resources at bus $b$}}
\item[\textcolor{black}{$\overline{G}_{b}^q/\underline{G}_{b}^q$}]{\textcolor{black}{Maximum/minimum reactive power output of controllable generation resources at bus $b$}}
\item[\textcolor{black}{$K$}]{\textcolor{black}{Power factor}}
\item[\textcolor{black}{$\overline{\mathcal{P}}_{t,b}^{\alpha\beta}$}]{\textcolor{black}{Default transition probability from state $\beta$ to state $\alpha$ for the TCL ensemble at bus $b$ during time interval $t$}}
\item[\textcolor{black}{$R_{l}$}]{\textcolor{black}{Resistance of line $l$}}
\item[\textcolor{black}{$X_{l}$}]{\textcolor{black}{Reactance of line $l$}}
\item[\textcolor{black}{$\overline{U}_{b}/\underline{U}_{b}$}]{\textcolor{black}{Maximum/minimum limit on the voltage squared at bus $b$}}
\item[\textcolor{black}{$\overline{V}_{b}/\underline{V}_{b}$}]{\textcolor{black}{Maximum/minimum limit on the voltage at bus $b$}}
\item[\textcolor{black}{$V_0$}]{\textcolor{black}{Base voltage squared at root-bus (substation)}}
\item[\textcolor{black}{$\delta$}]{\textcolor{black}{Exogenous step-scaling parameter}}
\item[\textcolor{black}{$\epsilon_{t,b}^{p/q}$}]{\textcolor{black}{Active/reactive forecast error}}
\item[\textcolor{black}{$\Tilde{\epsilon}_{t}^{p/q}$}]{\textcolor{black}{Aggregated active/reactive forecast error}}
\item[\textcolor{black}{$\zeta$}]{\textcolor{black}{Tolerance for the termination of the algorithm}}
\item[\textcolor{black}{$\eta_{g}/\eta_{v}$}]{\textcolor{black}{Violation tolerance on chance constraints}}
\item[\textcolor{black}{$\Lambda_{t}$}]{\textcolor{black}{Parameter to monetize active power losses}}
\end{IEEEdescription}

\subsection{Modeling a TCL Ensemble}
We assume that all TCLs which are co-located at the same bus of the distribution system are organized in one TCL ensemble and is operated by one aggregator. Each ensemble is assumed to have a sufficiently large (infinite) number of TCLs. Under this assumption, one can represent each TCL ensemble as a discrete-time and discrete-space MDP and consider that it is capable of maintaining scheduled injections with the distribution system (i.e. there are no fluctuations). The aggregator controls the TCL ensemble by optimizing its transition from one aggregated state to another across the optimization horizon. Similarly to our previous work in \cite{TCL_networks_Misha}, we use the MDP framework to build the following model for each TCL ensemble at bus $b$:
\allowdisplaybreaks
\begin{align}
&\underset{\substack{\rho,\mathcal{P}, p, q}}{\text{min}}\sum_{t \in \mathcal{T}} O_{b,t}^{A} := \mathbb{E}_{\rho}
\sum_{t \in \mathcal{T}} \! \sum_{\alpha \in \mathcal{A}} \! \left(-U_{t+1,b}^{\alpha} + \! \textcolor{black}{\sum_{\beta \in \mathcal{A}}} \gamma_{t,b}^{\alpha\beta} \log\! \frac{\mathcal{P}_{t,b}^{\alpha\beta}}{\overline{\mathcal{P}}_{t,b}^{\alpha\beta}}\right) 
\label{MDP:obj}
\\
&\rho_{t+1,b}^{\alpha} = \sum_{\beta \in \mathcal{A}} \mathcal{P}_{t,b}^{\alpha\beta} \rho_{t,b}^{\beta}, \quad \forall \alpha \in \mathcal{A}, t \in \mathcal{T} 
\label{MDP_evol} \\
& p_{t,b}= \sum_{\alpha \in \mathcal{A}} p_b^{\alpha} \rho_{t,b}^{\alpha}, \quad \forall t \in \mathcal{T}, b \in \mathcal{N} \label{mdp_injP1} \\
& q_{t,b} = \sum_{\alpha \in \mathcal{A}} q_b^{\alpha} \rho_{t,b}^{\alpha} , \quad \forall t \in \mathcal{T}, b \in \mathcal{N} \label{mdp_injQ1} \\
&\sum_{\alpha \in \mathcal{A}} \mathcal{P}_{t,b}^{\alpha \beta} = 1, \quad \forall t \in \mathcal{T}, \beta \in \mathcal{A}, b \in \mathcal{N}^{\mathcal{T}} \label{mdp_integrality} 
\end{align}
\allowdisplaybreaks[0]
where $\rho_{t,b}^{\alpha} \geq 0$ and $\rho_{t,b}^{\beta} \geq 0$ are decision variables that characterize the probability that TCLs at bus $b$ are operated in states $\alpha$ and $\beta$ and have active power consumptions modeled by parameters $p^{\alpha}_{t,b}$ and $p^{\beta}_{t,b}$, respectively. \textcolor{black}{The set of states for each TCL ensemble is denoted as $\mathcal{A}$ and individual states $\alpha$ and $\beta$ are $\alpha, \beta \in \mathcal{A}$.\footnote{Note that the ensemble can remain in the same state at time  $t$ and $t+1$. In this case state $\beta$ is such that $\beta = \alpha \in \cal{A}$.}
These states are obtained by discretizing the range of power consumption for each TCL ensemble given the operating range of each TCL (see Fig.~\ref{mdp:states}).} Decision variables $\rho_{t,b}^{\alpha}$ and $\rho_{t,b}^{\beta}$ are related via the transition probability $\mathcal{P}_{t,b}^{\alpha\beta}$ that characterizes the probability of the transition of TCLs at bus $b$ from state $\beta$ at time $t$ to state $\alpha$ at time $t+1$. On the other hand, parameter vector $\mathcal{\overline{P}}_{t,b}^{\alpha\beta}$ represents the default transition probability, i.e. internal dynamics of the TCL ensemble without actions of the aggregator. In practice, one can dynamically estimate $\mathcal{\overline{P}}_{t,b}^{\alpha\beta}$ from historical observations using reinforcement learning \cite{7401112}. 

\textcolor{black}{Eq.~\eqref{MDP:obj} represents the objective function of the aggregator that controls the TCL ensemble and aims to maximize the expected utility of the aggregator ($U_{t+1,b}^{\alpha}$) and to minimize the discomfort cost for the TCL ensemble, which is computed using the exogenous cost penalty ($\gamma^{\alpha \beta}_{t,b}$) and the Kullback-Leibler (KL) distance to penalize the difference between the transition decisions made by the aggregator ($\mathcal{P}_{t,b}^{\alpha\beta}$) and the default transitions of the TCL ensemble ($\mathcal{\overline{P}}_{t,b}^{\alpha \beta}$).} The choice of Kullback-Leibler distance for the penalty cost is motivated by its wide use for modeling randomness of discrete and continuous time-series. Other penalty functions can also be used instead. Eq.~\eqref{MDP_evol} describes the temporal evolution of the TCL ensemble, where initial conditions over the course of the optimization horizon are given by the decision of the aggregator during the previous optimization horizon. Eq.~\eqref{mdp_injP1}-\eqref{mdp_injQ1} computes the expected active and reactive power injections of the TCL ensemble to the distribution system. Eq.~\eqref{mdp_integrality} imposes the integrality constraint on the transition decisions optimized by the aggregator such that their total probability is equal to one. 

The optimization in Eq.~\eqref{MDP:obj}-\eqref{mdp_integrality} can be solved using dynamic programming that facilitates scalability of our approach and the ability to solve a large number of such optimizations, one for each TCL ensemble, in parallel. This property is particularly helpful when one deals with a large penetration of TCL ensembles anticipated in distribution systems of the future. Thus, in the following, we use a backward-forward algorithm to solve Eq.~\eqref{MDP:obj}-\eqref{mdp_integrality}. This algorithm is an iterative, two-step procedure that is commonly used for inferring probabilities of unknown state probabilities for Markov processes. We customize this procedure to find the optimal TCL transitions ($\mathcal{P}_{t,b}^{\alpha\beta}$) as further described in Appendix~\ref{sec:algoirthm}. Note that the optimization in Eq.~\eqref{MDP:obj}-\eqref{mdp_integrality} can be represented as a Linearly Solvable MDP (LS-MDP) \cite{LSMDP_todorov}, if $\gamma^{\alpha\beta}_{t,b}=0$, i.e., not state-dependent. Such LS-MDP problems can be solved analytically, i.e. without relying on iterative solution techniques \cite{LSMDP_Dvijotham,LSMDP_book,LSMDP_Meyn}, which can be exploited in  online dispatch applications.

\subsection{Chance Constrained Optimal Power Flow}
Proliferation of DERs imposes uncertainty on the nodal power injections (e.g. due to the solar irradiance). Unlike TCL ensembles, this uncertainty can be accurately parameterized using standard probability distributions and thus endogenously modeled in decision support tools used by the DSO in a computationally tractable manner. We therefore formulate a CC-OPF that takes the DSO perspective and seeks the least-cost strategy to operate the distribution system given its technical limits and PV uncertainty\footnote{ Additional uncertainty may also arise from the TCL ensembles due to a large but finite number of TCL users. However, as per the law of large numbers, these fluctuations scale as $\sim 1/\sqrt{N}$, where $N$ is a number of users, and are thus significantly smaller than $O(1)$ fluctuations of the PV resources.}.  
  
\subsubsection{Deterministic OPF} The CC-OPF is built based on the following deterministic OPF model that considers AC power flows using the \textit{LinDistFlow} model \cite{Lindistflow} and disregards the PV uncertainty:
  \begin{flalign}
&\underset{\substack{p^g,q^g,p^c,\\q^c,u,f^p,f^q}}{\text{min}}\sum_{t \in \mathcal{T}} \sum_{l\in\mathcal{E}} R_{l} \frac{f_{t,l}^{p^2} + f_{t,l}^{q^2}}{V_0^2} \label{OPF:obj} \\
&f_{t,l(b)}^{p} \!+ \!p_{t,b}^g\! +\! p_{t,b}^{\mathcal{PV}} \!=\! p_{t,b}^c\! +\! p_{t,b} +\!\! \sum_{l|o(l)=b} \!\!f_{t,l}^p, \forall t \!\in\! \mathcal{T}, \!b\! \in\! \mathcal{N} \label{opf_nodep}\\
&f_{t,l(b)}^{q} \!+ \!q_{t,b}^g\! +\! q_{t,b}^{\mathcal{PV}}\! =\! q_{t,b}^c\! + \!q_{t,b}\! +\!\! \sum_{l|o(l)=b} \!\!f_{t,l}^q, \forall t \in \mathcal{T}, \!b\! \in\! \mathcal{N} \label{opf_nodeq}\\
&u_{t,r(l)} = u_{t,o(l)} - 2(R_{l}f_{t,l}^{p} + X_{l}f_{t,l}^{q}),\ \forall t \in \mathcal{T},\ l \in \mathcal{E} \label{opf_vol}\\
&\underline{G}_{b}^{p} \leq {p_{t,b}^g} \leq \overline{G}_{b}^{p} \label{opf_gp},\ \forall t \in \mathcal{T},\ b\in \mathcal{N}^{\mathcal{G}} \\
&\underline{G}_{b}^{q} \leq {q_{t,b}^g} \leq \overline{G}_b^{q} \label{opf_gq},\ \forall t \in \mathcal{T},\ b\in \mathcal{N}^{\mathcal{G}} \\
&\underline{U}_b \leq u_{t,b} \leq \overline{U}_b \label{opf_vol_lim},\ \forall t \in \mathcal{T}, b \in \mathcal{N} 
\end{flalign}
Eq.~\eqref{OPF:obj} minimizes the active power losses in the distribution system. Note that the proposed formulation and algorithm can accommodate other choices of the objective function (e.g. cost-minimization). Eq.~\eqref{opf_nodep}-\eqref{opf_vol} are nodal active and reactive power balances as in the \textit{LinDistFlow} model \cite{Lindistflow}. \textcolor{black}{Although Eq.~\eqref{opf_nodep}-\eqref{opf_vol} neglect the effect of power losses, the objective function in Eq.~\eqref{OPF:obj} can still be formulated in a loss-minimization manner based on the active and reactive power flows $f_{t,l}^{p}$ and $f_{t,l}^{q}$ provided by the \textit{LinDistFlow} model.} In Eq.~\eqref{opf_nodep}-\eqref{opf_nodeq}, $p_{t,b}$ and $q_{t,b}$ are parameterized and obtained from the MDP optimization in Eq.~\eqref{MDP:obj}-\eqref{mdp_integrality}. Eq.~\eqref{opf_gp}-\eqref{opf_gq} enforce the minimum and maximum limits on the active and reactive power output of controllable generation resources. Eq.~\eqref{opf_vol_lim} limits voltage magnitudes squared within their minimum and maximum values. 

\subsubsection{PV Uncertainty} The uncertain PV output at every bus $b$ and time interval $t$ is defined as $\boldsymbol{p}_{t,b}^{\mathcal{PV}} = p_{t,b}^{\mathcal{PV}} {-} \epsilon_{t,b}^p $, where $p_{t,b}^{\mathcal{PV}}$ is the forecast value and $\epsilon_{t,b}^p$ is a forecast error. We assume that this forecast error follows a zero-mean, normal distribution with variance $\sigma_{t,b}^2$, i.e. $\epsilon_{t,b}^p \sim N(0, \sigma_{t,b}^2)$. The forecast error in that form is commonly provided by forecast vendors (e.g. \cite{BACHER20091772}). Furthermore, the inaccuracy of assuming normally distributed forecast errors can be mitigated in the CC-OPF using data-robust approaches as in \cite{Robert_cdc_2018, CCopf_Lubin}. Since active power forecast errors also cause fluctuations of the reactive power, we assume that the latter errors are proportional, i.e. $\epsilon_{t,b}^q = \epsilon_{t,b}^p K$, where $K$ is a parameter computed for a given power factor.  To compensate for the forecast error and thus to maintain the generation-load balance, controllable generators operated by the DSO are assumed to adjust their output based on a proportional control law, \cite{CCopf_Bienstock, Robert_cdc_2018, CCopf_Lubin}. This control assumes that each controllable generator compensates a fraction of the aggregated forecast error by changing its real-time active ($\boldsymbol{p}_{t,b}^g$) and reactive ($\boldsymbol{q}_{t,b}^g$) power outputs around its generation setpoints $p_{t,b}^g$ and $q_{t,b}^g$ optimized for a given forecast based on optimized participation factors $\alpha_{t,b}$. This control is formalized as:
\begin{align}
&\boldsymbol{p}_{t,b}^g = p_{t,b}^g {+} \alpha_{t,b} \Tilde{\epsilon}_{t}^p,\ \ \ \forall t,\ b\in\mathcal{N}^{\mathcal{G}} \label{CC_pg}\\
&\boldsymbol{q}_{t,b}^g = q_{t,b}^g {+} \alpha_{t,b} \Tilde{\epsilon}_{t}^q,\ \ \ \forall t,\ b\in\mathcal{N}^{\mathcal{G}} \label{CC_qg}
\end{align}
where $\Tilde{\epsilon}_{t}^p$ = $\sum_{b\in \mathcal{N}^{\mathcal{PV}} } \epsilon_{t,b}^p$ and $\Tilde{\epsilon}_{t}^q$ = $\sum_{b\in \mathcal{N}^{\mathcal{PV}} } \epsilon_{t,b}^q$ are the aggregated forecast errors for every time interval $t$. Participation factors $\alpha_{t,b}$ are optimized to accommodate different technical and cost characteristics of controllable generators with the condition $\sum_{b\in \mathcal{N}^{\mathcal{G}}} \alpha_{t,b}$ = 1, $\forall t \in \mathcal{T}$, i.e. the total change in the output of controllable generators is equal to the aggregated forecast error.

Following the realization of $\epsilon_{t,b}^p$ and $\epsilon_{t,b}^q$, the real-time active ($\boldsymbol{f}_{t,l}^p$) and reactive ($\boldsymbol{f}_{t,l}^q$) power flows are modeled as:
\begin{align}
&\boldsymbol{f}_{t,l}^p = f_{t,l}^p + a_{l*}(\epsilon_{t}^p - \alpha_{t} \Tilde{\epsilon}_{t}^p),\ \ \ \forall t,\ \forall l \label{CC_fp}\\
&\boldsymbol{f}_{t,l}^q = f_{t,l}^q + a_{l*}(\epsilon_{t}^q - \alpha_{t} \Tilde{\epsilon}_{t}^q),\ \ \ \forall t,\ \forall l \label{CC_fq}
\end{align}
where $a_{l*}$ denotes the $l^{\text{th}}$ row of a matrix $A: |\mathcal{E}| \times |\mathcal{N}|$ with elements $a_{(lb)}$ which we define such that:
\[ a_{(lb)} =
 \begin{cases}
  1, &\text{if line $l$ is part of the path}\\ 
    &\text{from root to bus $b$} \qquad \ \ \ l\in \mathcal{E}, b\in \mathcal{N}\\
  0, &\text{otherwise}
 \end{cases}
\]
and $\alpha_t$ is a $|\mathcal{N}| \times 1$ vector with elements $\alpha_{t(b)}$ such that: 
\[ \alpha_{t(b)} =
 \begin{cases}
  \alpha_{t,b}, &\text{if} \ \ b\in \mathcal{N}^{\mathcal{G}} \\
  0, &\text{otherwise}. 
 \end{cases}
\]

Accordingly, one can use Eq.~\eqref{CC_fp}-\eqref{CC_fq} to derive the real-time voltage magnitudes squared ($\boldsymbol{u}_{t,b}$):
\begin{align}
\begin{split}
&\boldsymbol{u}_{t,r(l)} = u_{t,r(l)} - 2 a_{*r(l)}^{\top} [R A(\epsilon_{t}^p - \alpha_{t} \Tilde{\epsilon}_{t}^p) \\&+ X A(\epsilon_{t}^q - \alpha_{t} \Tilde{\epsilon}_{t}^q)], \quad \forall t \in \mathcal{T}, b \in \mathcal{N} \label{CC_u}
\end{split}
\end{align}
where $R$, $X$ are $|\mathcal{E}| \times |\mathcal{E}|$ matrices with diagonal entries consisting of the line resistances and reactances respectively: $R_{(ii)} = R_i, R_{(ij, i \neq j)} = 0$, $X$ in analogy.

\subsubsection{Formulation}
Using the results in Eqs.~\eqref{CC_pg}-\eqref{CC_u}, the deterministic OPF in Eq.~\eqref{OPF:obj}-\eqref{opf_vol_lim} can be converted into the following CC-OPF formulation that accounts for real-time quantities $\boldsymbol{p}_{t,b}^g$, $\boldsymbol{q}_{t,b}^g$, $\boldsymbol{f}_{t,l}^p$, $\boldsymbol{f}_{t,l}^q$ and $\boldsymbol{u}_{t,b}$ following realizations of $\epsilon_{t,b}^p$ and $\epsilon_{t,b}^q$:
\begin{align}
&\underset{\substack{p^g,q^g,\\p^c,q^c,u,f^p,\\f^q,\alpha}}{\min} \sum_{t \in \mathcal{T}} O^{D}_t = \mathbb{E}_{\epsilon^p, \epsilon^q} \bigg[
 \sum_{t \in \mathcal{T}} \sum_{l \in \mathcal{E}} R_{l} \frac{\boldsymbol{f}_{t,l}^{p^2} + \boldsymbol{f}_{t,l}^{q^2}}{V_0^2}
\bigg] \label{CC:obj} \\
&\text{Eq.~\eqref{opf_nodep}-\eqref{opf_vol}}, \label{CC:first_const}\\
&\text{Eq.~\eqref{CC_pg}-\eqref{CC_qg}, \eqref{CC_u} }\\
&\mathbb{P}( {\boldsymbol{p}_{t,b}^g} \leq \overline{G}_b^{p}) \geq (1 - \eta_{g}),\ \forall t \in \mathcal{T},\ b\in \mathcal{N}^{\mathcal{G}} \label{CC_gplim1}\\
&\mathbb{P}(\underline{G}_{b}^{p} \leq {\boldsymbol{p}_{t,b}^g}) \geq (1 - \eta_{g}),\ \forall t \in \mathcal{T},\ b\in \mathcal{N}^{\mathcal{G}} \label{CC_gplim2}\\
&\mathbb{P}({\boldsymbol{q}_{t,b}^g} \leq \overline{G}_b^{q}) \geq (1 - \eta_{g}),\ \forall t \in \mathcal{T}, b \in \mathcal{N}^{\mathcal{G}} \label{CC_gqlim1}\\
&\mathbb{P}(\underline{G}_{b}^{q} \leq {\boldsymbol{q}_{t,b}^g}) \geq (1 - \eta_{g}),\ \forall t \in \mathcal{T}, b \in \mathcal{N}^{\mathcal{G}} \label{CC_gqlim2}\\
&\mathbb{P}( {\boldsymbol{u}_{t,b}} \leq \overline{U}_b) \geq (1 - \eta_{v}),\forall t \in \mathcal{T}, b \in \mathcal{N} \label{CC_vlim1}\\
&\mathbb{P}(\underline{U}_{b} \leq {\boldsymbol{u}_{t,b}}) \geq (1 - \eta_{v}),\forall t \in \mathcal{T}, b \in \mathcal{N}, \label{CC_vlim2}
\end{align}
where $\eta_g$ and $\eta_v$ are exogenous parameters that define tolerance to constraint violations. Eq.~\eqref{CC_gplim1}-\eqref{CC_vlim2} are chance constraints on the power outputs of conventional generators and voltage magnitudes squared that replace deterministic constraints in Eq.~\eqref{opf_gp}-\eqref{opf_vol_lim}. Under the assumption that \mbox{$\eta_g, \eta_v < 0.5$}, one can recast Eqs.~\eqref{CC_gplim1}-\eqref{CC_vlim2} into SOC constraints that can in turn be solved efficiently using off-the-shelf solvers, \cite{CCopf_Bienstock, CCopf_Lubin}. The reformulation process for Eq.~\eqref{CC_gplim1}-\eqref{CC_vlim2} is shown in Appendix \ref{Appendix:conic_refor}. Accordingly, the expected value in the objective function given by Eq.~\eqref{CC:obj} is derived below as:
\begin{align}
&\sum_{t \in \mathcal{T}} O_t^{D} = \sum_{t\in\mathcal{T}} \sum_{l\in\mathcal{E}}\frac{R_{l}}{V_0^2}\big(\mathbb{E}[\boldsymbol{f}_{t,l}^{p^2}] + \mathbb{E}[\boldsymbol{f}_{t,l}^{q^2}]\big), \label{expected_ini}
\end{align}
where $\mathbb{E}[\boldsymbol{f}_{t,l}^{q^2}]$ and $\mathbb{E}[\boldsymbol{f}_{t,l}^{q^2}]$ are as follows:
\begin{align}
 \mathbb{E}[\boldsymbol{f}_{t,l}^{p^2}] &= \text{Var}(\boldsymbol{f}_{t,l}^p) + (f_{t,l}^{p}){^2} \nonumber \\
    &= \sum_{j\in\mathcal{N}}\!\! \big[ a_{lj} ( \text{Var} (\epsilon_{t,j}^p) - \alpha_{t,j}^2 \text{Var}(\Tilde{\epsilon}_{t}^p))\big] + (f_{t,l}^{p}){^2} \label{eq:_obj_exp1ref} \\
 \mathbb{E}[\boldsymbol{f}_{t,l}^{q^2}] &= \text{Var}(\boldsymbol{f}_{t,l}^q) + (f_{t,l}^{q}){^2} \nonumber \\
    &=\! \sum_{j\in\mathcal{N}}\!\!\big[ a_{lj} ( \text{Var} (\epsilon_{t,j}^q) - \alpha_{t,j}^2 \text{Var}(\Tilde{\epsilon}_{t}^q))\big] + (f_{t,l}^{q}){^2} \label{eq:_obj_exp2ref}
\end{align}
 Given Eq.~\eqref{eq:_obj_exp1ref}-\eqref{eq:_obj_exp2ref}, the right hand-side of the original objective function in Eq.~\eqref{expected_ini} can be re-written as:
\begin{align}
& \sum_{t \in \mathcal{T}} O_t^{D} =\!\! \sum_{t\in\mathcal{T}} \sum_{l\in\mathcal{E}}\!\! \frac{R_{l}}{V_0^2} \bigg[ \sum_{j\in\mathcal{N}} \big[ a_{lj} ( \text{Var} (\epsilon_{t,j}^p) -\!\! \alpha_{t,j}^2 \text{Var}(\Tilde{\epsilon}_{t}^p))\big] \!+\! f_{t,l}^{p^2} \notag \\&\!\! \qquad + \sum_{j\in\mathcal{N}}\!\!\big[ a_{lj} ( \text{Var} (\epsilon_{t,j}^q) \!-\! \alpha_{t,j}^2 \text{Var}(\Tilde{\epsilon}_{t}^q))\big] \!+\! f_{t,l}^{q^2} \bigg]. \label{eq:objective_function_analytical} 
\end{align}
Thus, Eq.~\eqref{eq:objective_function_analytical} is a quadratic, deterministic equivalent of the original objective function permissible for off-the-shelf solvers. 

\textit{\textcolor{black}{Remark 1:}} \textcolor{black}{While the Gaussian assumption to represent the PV uncertainty is sufficient for the needs of this paper, it is not generally restrictive. As shown in \cite{Robert_cdc_2018, Roald_DRCC, Dvorkin_uncertainty_sets}, non-Gaussian distributions or their approximation via a Gaussian mixture can be used to reformulate the chance constraints without increasing computational complexity of the model. Such reformulations tend to yield a more robust, but expensive solution.}

\textit{Remark \textcolor{black}{2}:} Note that the CC-OPF in Eq.\eqref{CC:obj}-\eqref{CC_vlim2} does not impose power flow limits on $\boldsymbol{f}_{t,l}^p$ and $\boldsymbol{f}_{t,l}^q$, because distribution systems are typically voltage-constrained and power flow limits can be disregarded. In \cite{line_miles_2018}, we describe an approach to enforce chance-constrained apparent power limits.

\textit{\textcolor{black}{Remark 3:}} \textcolor{black}{The PV uncertainty is modeled with individual chance constraints as given by Eq.~\eqref{CC_gplim1}-\eqref{CC_vlim2}, which are computationally tractable. Alternatively, one can consider the joint chance constraint over Eq.~\eqref{CC_gplim1}-\eqref{CC_vlim2}, but such a formulation would be computationally unbearable \cite{CC_joint}. Since replacing the joint chance constraint with individual chance constraints may lead to a conservative solution, especially for large networks, one needs to tune the value of parameters $\eta_g$ and $\eta_v$.}

\subsection{Integrated Optimization Problem}
Based on the models in Eq.~\eqref{MDP:obj}-\eqref{mdp_integrality} and in Eq.~\eqref{CC:obj}-\eqref{CC_vlim2}, the integrated optimization problem that includes MDP and CC-OPF is formulated as follows: 
\begin{align}
\begin{split}
&\underset{\substack{\rho,\mathcal{P},u,p,q\\f^p,f^q,\\p^c,q^c, p^g,q^g}}{\text{min}}
\sum_{t \in \mathcal{T}} \Bigg[ \sum_{b \in \mathcal{N}^{\mathcal{T}}} O^A_{b,t} + \Lambda_t O^D_t \Bigg] \label{obj}
\end{split}\\
& \text{Eq.}~\eqref{MDP_evol}-\eqref{mdp_integrality} \label{int_mdp}\\
& \text{Eq.}~\eqref{CC:first_const}-\eqref{CC_vlim2}, \label{int_opf}
\end{align}
where parameter $\Lambda_t$ is a tariff that monetizes the active power losses to make them comparable to the MDP objective function. The optimization in Eq.~\eqref{obj}-\eqref{int_opf} cannot be solved efficiently using existing dynamic programming methods and off-the-shelf solvers. This motivates the solution technique described in Section~\ref{Sec:ST}.

\section{Solution Technique} \label{Sec:ST}

\textcolor{black}{To solve the integrated problem in \eqref{obj}-\eqref{int_opf}, we propose a Spatio-Temporal Dual Decomposition (ST-D2) algorithm that exploits two ideas.} First, we seek the consensus between the MDP and CC-OPF and thus use a dual decomposition of the original problem. Second, we decouple some spatially- and temporally-independent decision variables. The spatial separation is applied because each TCL ensemble is located at a unique bus and therefore can be optimized separately. In this case, the MDP optimization for each TCL ensemble is performed over the entire optimization horizon to capture inter-temporal constraints on each TCL ensemble. On the other hand, the CC-OPF decisions are temporally separable since controllable generators located in the distribution system typically do not have such inter-temporal constraints as ramping rate and minimum up/down time limits, which are customary for transmission systems. Therefore, the CC-OPF can be solved separately for each time interval $t$, see \cite{CCopf_Emiliono} for a time-decoupled OPF example. \textcolor{black}{In the presence of temporally coupled resources, e.g. energy storage systems, which introduce the inter-temporal constraints, the CC-OPF can be solved over the entire optimization horizon. This will increase computing times, but is still computationally tractable as shown in \cite{Robert_cdc_2018}.}

 The proposed ST-D2 algorithm iterates as illustrated in Fig.~\ref{fig:std2} and each step is further itemized below, where $\nu$ is an iteration counter:
 
\begin{figure}[!t]
\centering
\tikzstyle{decision} = [diamond, draw, fill=white!10, 
  text width=7.5em, text badly centered, node distance=1.50cm, inner sep=-8pt]
\tikzstyle{block} = [rectangle, draw, fill=white!10, 
  text width=15em, text centered, rounded corners, node distance=1.25cm]
\tikzstyle{line} = [draw, -latex']
\tikzstyle{cloud} = [draw, ellipse,fill=red!20, node distance=2cm,
  minimum height=2em]
  
\begin{tikzpicture}[node distance = 1.25cm, auto]
  \node [block] (init) {Initialize model};
  \node [block, below of=init, node distance=1cm] (mdp_step1) {MDP optimization (Step 1)};
  \node [block, below of=mdp_step1] (ccopf_step2) {CC OPF optimization (Step 2)};
  \node [block, below of=ccopf_step2, node distance=1.25cm] (Lagrange_step3) {Update Lagrange multiplies (Step 3)};
  \node [decision, below of=Lagrange_step3, node distance=2.25cm] (decide) {Lagrange multipliers converged?};
  \node [block, below of=decide, node distance=2.25cm] (stop) {Stop};
  \coordinate[left of=decide, xshift=-2.0cm] (c1);
  \coordinate[left of=mdp_step1, xshift=-2.0cm] (c2);
  \path [line] (init) -- (mdp_step1);
  \path [line] (mdp_step1) -- node {$p_{t,b}^{(\nu)}, q_{t,b}^{(\nu)}$} (ccopf_step2);
  \path [line] (ccopf_step2) -- node{$\lambda_{t,b}^{p(\nu)}$, $\lambda_{t,b}^{q(\nu)}$}(Lagrange_step3);
  \path [line] (Lagrange_step3) -- (decide);
  \path [line] (decide) -- node {Yes}(stop);
  \draw [-,thick, line width=0.2mm] (decide) -- (c1) node [midway] {No};
  \draw [-,thick, line width=0.2mm] (c1) -- (c2) node [midway, rotate=90, xshift=1.5cm,  yshift=+0.5cm] {$\lambda_{t,b}^{p(\nu+1)}, \lambda_{t,b}^{q(\nu+1)}$};
  \draw [->,thick, line width=0.2mm] (c2) -- (mdp_step1);
\end{tikzpicture}
\vspace{5pt}
\caption{Spatio-Temporal Dual Decomposition (ST-D2) Algorithm. }
\label{fig:std2}
\end{figure}
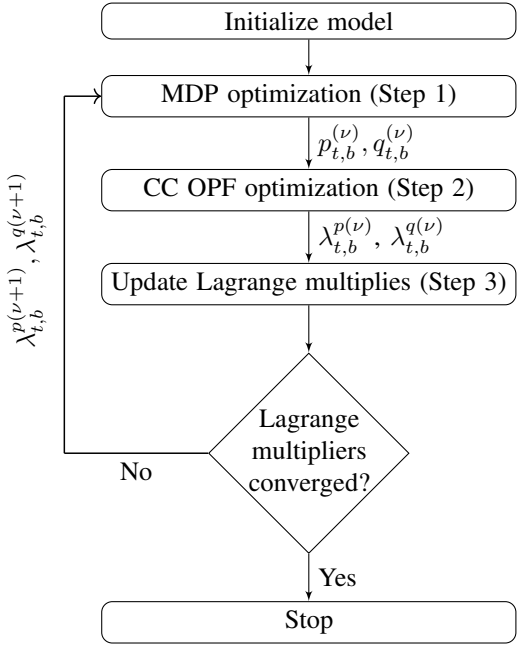

\begin{enumerate}
  \item Solve the MDP for each TCL ensemble:
  \begin{align*}
\forall b \in \mathcal{N}^{\mathcal{T}}: \quad & \underset{\substack{\rho,\mathcal{P}}}{\text{min}}
\sum_{t \in \mathcal{T}} O_{b,t}^{A(\nu)} \\
\hspace{-5mm} &\text{Eq.}~\eqref{MDP_evol}-\eqref{mdp_integrality} \\ 
\hspace{-5mm} &U_{t+1,b}^{\alpha(\nu)}\!\! =\!\! U_{t,b}^{\alpha(\nu)} \!+\! \lambda_{t,b}^{p(\nu)} p_b^{\alpha}\! + \!\lambda_{t,b}^{q(\nu)} q_b^{\alpha}, \nonumber \\ & \quad\quad\quad\quad\quad\quad\quad\quad\quad\quad\quad\quad \forall \alpha \!\in \!\mathcal{A},\! t\! \in \mathcal{T}
\end{align*}
where $\lambda_{t,b}^{p(\nu)}$ and $\lambda_{t,b}^{q(\nu)}$ are the Lagrange multipliers of Eq.~\eqref{mdp_injP1} and \eqref{mdp_injQ1}, respectively, obtained at the previous iteration of the ST-D2 algorithm. 
Hence, \mbox{$\lambda_{t,b}^{p(\nu=1)}=\lambda_{t,b}^{q(\nu=1)}=0$} during the first iteration.
  \item Solve the CC-OPF problem, where each TCL ensemble is parameterized using the values of Lagrange multipliers $\lambda_{t,b}^{p(\nu)}$:
\begin{align*}
\begin{split}
\forall t \in \mathcal{T}: & \underset{\substack{p^g,q^g,p^c,\\q^c,u,f^p,f^q}} \min \sum_{t \in \mathcal{T}} \Lambda_t O^{D(\nu)}_t \\& - \sum_{b\in\mathcal{N}}(\lambda_{t,b}^{p(\nu)}p_{t,b}^{(\nu)} + \lambda_{t,b}^{q(\nu)}q_{t,b}^{(\nu)}) 
\end{split}\\
&\text{Eq.}~\eqref{CC:first_const}-\eqref{CC_vlim2}, 
\end{align*}
where the CC-OPF problems for all time intervals are solved in parallel.
  \item Update the Lagrange multipliers:
\begin{align*}
&\lambda_{t,b}^{p(\nu+1)} \leftarrow \lambda_{t,b}^{p(\nu)} + \delta\bigg(\sum_{\alpha \in \mathcal{A}}p_b^{\alpha }\rho_{t,b}^{\alpha(\nu)} - p_{t,b}^{(\nu)}\bigg)\\
&\lambda_{t,b}^{q(\nu+1)} \leftarrow \lambda_{t,b}^{} + \delta\bigg(\sum_{\alpha \in \mathcal{A}}q_b^{\alpha(\nu)}\rho_{t,b}^{\alpha(\nu)} - q^{(\nu)}_{t,b}\bigg)
\end{align*}
where $\delta$ is an exogenous parameter that can be tuned to improve computational performance \cite{ADMM}. 
\end{enumerate}
These iterations continue until $\lambda_{t,b}^{p(\nu)}$ and $\lambda_{t,b}^{q(\nu)}$ converge with a given termination tolerance (\textcolor{black}{$\zeta$}).

\textit{\textcolor{black}{Remark 4:}} \textcolor{black}{The proposed ST-D2 algorithm is based on the dual decomposition and, therefore, it inherits algorithmic properties of the dual decomposition, including convergence properties and the ability to deal with non-convex decisions, e.g. \cite{6760235}. Furthermore, it can be extended to the Alternating Direction Method of Multipliers (ADMM), which has gained attention in distributed power grid applications \cite{Petr_ADMM, ADMM_dist_EMM}, by adding the Augmented Lagrangian terms to the objective function of the integrated problem. As pointed out in \cite{Boyd}, the ADMM algorithm blends the benefits of the  dual decomposition and Augmented Lagrangian methods.}

\section{Case Study} \label{Sec:Case_study}

\begin{figure}[!b]
  \centering
\begin{forest}for tree={circle,draw,minimum size=1.5em, 
      inner sep=1pt,l sep=0.1cm}
 [1,fill=green,before computing xy={l=10mm,s=0mm}
 [2,fill=yellow
 [3,fill=yellow,before computing xy={l=10mm,s=0mm}[4,before computing xy={l=10mm,s=0mm}[5[6,fill=yellow[7,before computing xy={l=0mm,s=-10mm}[8,before computing xy={l=-10mm,s=0mm}[9,before computing xy={l=-10mm,s=0mm}[10,before computing xy={l=0mm,s=-10mm}[11,before computing xy={l=0mm,s=-10mm}[12[13[14,fill=orange[15,before computing xy={l=0mm,s=10mm}[16,before computing xy={l=0mm,s=10mm}[17,fill=cyan,before computing xy={l=0mm,s=10mm}[18,fill=yellow,before computing xy={l=0mm,s=10mm}]]]]]]]]]]]][26,fill=cyan,fill=cyan,before computing xy={l=0mm,s=10mm}[27,before computing xy={l=-10mm,s=0mm}[28,before computing xy={l=-10mm,s=0mm}[29,before computing xy={l=-10mm,s=0mm}[30,before computing xy={l=0mm,s=10mm}[31,before computing xy={l=0mm,s=10mm}[32,fill=yellow,before computing xy={l=0mm,s=10mm}[33,before computing xy={l=10mm,s=0mm}]]]]]]]]]]][23,fill=cyan,before computing xy={l=0mm,s=-10mm}[24,before computing xy={l=0mm,s=-10mm}[25,fill=yellow,before computing xy={l=0mm,s=-10mm}]]]][19,before computing xy={l=0mm,s=10mm}[20,fill=cyan,before computing xy={l=0mm,s=10mm}[21,fill=yellow,before computing xy={l=0mm,s=10mm}[22,before computing xy={l=0mm,s=10mm}]]]]]]
\end{forest}
\vspace{3pt}
\caption{A schematic representation of the IEEE 33-bus distribution test system \cite{33_bus}, where the root bus is denoted in green (\#~1), bus with distributed generator in orange (\#~14), buses with PV resources in yellow (\#~2, 3, 6, 18, 21, 25, 32), and buses with TCL ensembles in blue (\#~17, 20, 23, 26).}
\label{dist_network}
\end{figure}
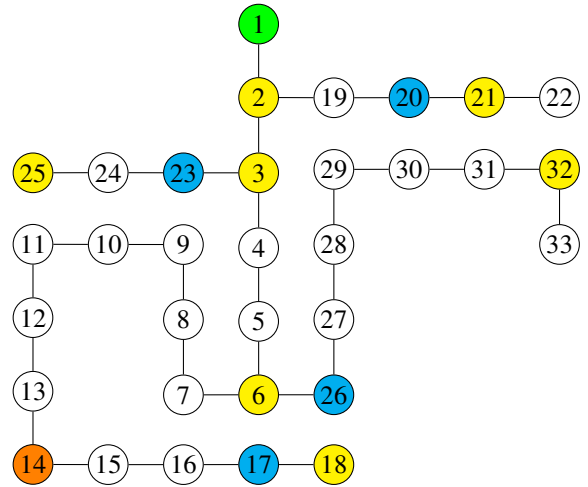

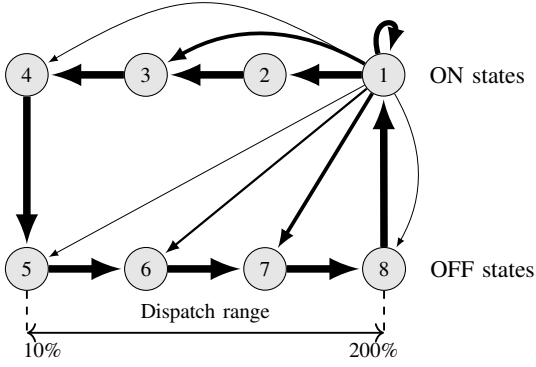
\begin{figure}
\centering
  \begin{tikzpicture}[font=\footnotesize]
  \node[state,fill=gray!20!white,minimum size=0.5cm] (s1) {1};
  \node[state,fill=gray!20!white,
     left=1cm of s1,minimum size=0.5cm] (s2) {2};
  \node[state,fill=gray!20!white,
     left=1cm of s2,minimum size=0.5cm] (s3) {3};
  \node[state,fill=gray!20!white,
     left=1cm of s3,minimum size=0.5cm] (s4) {4};
   \node[state,fill=gray!20!white,
     below=2cm of s4,minimum size=0.5cm] (s5) {5};
  \node[state,fill=gray!20!white,
     right=1cm of s5,minimum size=0.5cm] (s6) {6};
  \node[state,fill=gray!20!white,
     right=1cm of s6,minimum size=0.5cm] (s7) {7};
  \node[state,fill=gray!20!white,
     right=1cm of s7,minimum size=0.5cm] (s8) {8};
  \node[right=0.2cm of s1,font=\small,text width=1.5cm] {ON states};
  \node[right=0.2cm of s8,font=\small,text width=1.5cm] {OFF states};
  \coordinate[below of=s5, yshift=0.15cm] (c1);
  \coordinate[below of=s8, yshift=0.15cm] (c2);
  \draw[>=latex,every loop,fill=black!70,
    draw=black!70,
     auto=right,
     line width=0.5mm]
    (s1) edge[line width=1mm] (s2)
    (s2) edge[line width=1mm] (s3)
    (s3) edge[line width=1mm] (s4)
    (s4) edge[line width=1mm] (s5)
    (s5) edge[line width=1mm] (s6)
    (s6) edge[line width=1mm] (s7)
    (s7) edge[line width=1mm] (s8)
    (s8) edge[line width=1mm] (s1)
    (s1) edge[loop above,line width=0.7mm]  (s1)
    (s1) edge[bend right, auto=right] (s3)
    (s1) edge[bend right, auto=right,looseness=1.3,line width=0.1mm] (s4)
    (s1) edge[line width=0.1mm] (s5)
    (s1) edge[line width=0.3mm] (s6)
    (s1) edge[line width=0.5mm] (s7)
    (s1) edge[bend left, auto=left,line width=0.1mm] (s8);
  \draw [-,thick,dashed, line width=0.25mm] (s5) -- (c1) node [near start] {};
  \draw [-,thick,dashed, line width=0.25mm] (s8) -- (c2) node [near start] {};
  \draw [{<[scale=1.1]}-{>[scale=1.1]},thick, line width=0.25mm,anchor=south] (c1) -- (c2) node [midway] {Dispatch range} node [very near start,anchor=north east] {10\%} node [very near end,anchor=north west] {200\%};
  
  \end{tikzpicture}
  \vspace{4pt}
  \caption{A MDP representation of the TCL ensemble with 8 states displaying all possible transitions from state $1$. The active and reactive power consumptions at each state are obtained as uniform intervals within 10\% - 200\% of the average load of the TCL ensemble.}
  \label{mdp:states}
\end{figure}

The case study uses the IEEE 33-bus distribution system \cite{33_bus}, as shown in Fig.~\ref{dist_network}, where the root bus of the distribution system is connected to the transmission network. The root bus is considered as an infinite source of power supply. One controllable distributed generator with the maximum capacity of 30 kW is placed at bus \#~14. We consider that the power supply cost from the root bus and from the distributed generator is equal and set $\Lambda_t=\$10$/kWh. Seven PV resources, with the rated capacity of 1.5 kW, are installed at buses \#~2, 3, 6, 18, 21, 25 and 32 and produce at zero cost. The forecast error of each PV resource is zero-mean and its standard deviation is set to 30\% of the forecast output unless stated otherwise. The loads at buses \#~17, 20, 23 and 26 are replaced with TCL ensembles of the equivalent capacity. Each ensemble is discretized in 8 states, as shown in Fig.~\ref{mdp:states}, with the default transition probabilities ($\overline{\mathcal{P}}_{t,b}$) between the states as shown in Table~\ref{Table:mdp_states}. Each TCL ensemble can be dispatched in the the range of 10\% - 200\% of its average load. To assess the impact of TCL users' comfort, the cost penalty ($\gamma_{t,b}^{\alpha\beta}$) is considered for two cases. \textcolor{black}{The first case, referred to in the following as the \emph{uniform cost penalty} case, assumes the same penalty for each possible transition shown in Fig.~\ref{mdp:states}, i.e. $\gamma_{t,b}^{\alpha\beta}=1\$$. The second case, referred to in the following as the \emph{non-uniform cost penalty} case, differentiates between the transitions along the cycle (e.g., $1\rightarrow2\rightarrow3\rightarrow4\rightarrow5\rightarrow6\rightarrow7\rightarrow8\rightarrow1$) and other transitions (e.g., $1\rightarrow3, 1\rightarrow4,1\rightarrow5,1\rightarrow6,1\rightarrow7,1\rightarrow8$), where the former transitions are penalized using $\gamma_{t,b}^{\alpha\beta}=1\$$ and the latter transitions are penalized with $\gamma_{t,b}^{\alpha\beta}=10\$$. This differentiation allows to put a higher cost penalty on transitions with a larger power change that are more disruptive for TCL users' comfort.} The optimization horizon consists of 24 hourly time intervals. For the sake of simplicity, we assume that $\eta_v=\eta_g$.

All simulations are performed in Julia JuMP \cite{jump} using the Ipopt solver on an Intel Core i5 1.6 GHz processor with 4~GB of RAM. The value of \textcolor{black}{$\zeta$} is set to 0.0001. The code and input data used in this paper are available in \cite{code}. 

\begin{table}[!t]
\centering
\scriptsize
\caption{Default transition probabilities of each TCL ensemble between eight states in Figure~\ref{mdp:states}.}
\vspace{3pt}
 \begin{tabular}{c|cccccccc}
  \hline
  State & 1 & 2 & 3 & 4 & 5 & 6 & 7 & 8 \\ \hline
  1   & 0.2 & 0.5 & 0.1 & 0.03 & 0.02 & 0.03 & 0.1 & 0.02 \\ 
  2   & 0.02 & 0.2 & 0.5 & 0.1 & 0.03 & 0.02 & 0.03 & 0.1 \\ 
  3   & 0.1 & 0.02 & 0.2 & 0.5 & 0.1 & 0.03 & 0.02 & 0.03 \\ 
  4   & 0.03 & 0.1 & 0.02 & 0.2 & 0.5 & 0.1 & 0.03 & 0.02 \\ 
  5   & 0.02 & 0.03 & 0.1 & 0.02 & 0.2 & 0.5 & 0.1 & 0.03 \\ 
  6   & 0.03 & 0.02 & 0.03 & 0.1 & 0.02 & 0.2 & 0.5 & 0.1 \\ 
  7   & 0.1 & 0.03 & 0.02 & 0.03 & 0.1 & 0.02 & 0.2 & 0.5 \\ 
  8   & 0.5 & 0.1 & 0.03 & 0.02 & 0.03 & 0.1 & 0.02 & 0.2\\ 
  \hline
 \end{tabular}
 \label{Table:mdp_states}
\end{table}

\subsection{Computational Performance}
In the following numerical experiments, the proposed ST-D2 algorithm converges in 4-7 iterations. There is no explicit correlation observed between the complexity of the problem (number of TCL ensembles considered) and the number of iterations and computing times required for convergence. For example, the most complex instance with four TCL ensembles is solved in \textcolor{black}{189.90} seconds. Table~\ref{table:cp} itemizes this computing time for the MDP optimization (Step 1) and CC-OPF optimization (Step 2). Wihtin these four iterations, only \textcolor{black}{19.92} seconds ($\approx \textcolor{black}{10.4\%}$ of the total computing time) is spent on the MDP optimization in Step 1, while the rest of the time is incurred by the CC-OPF optimization in Step 2. These results demonstrate that the MDP optimization has \textcolor{black}{one-tenth} bearing on the overall complexity of the ST-D2 algorithm relative to the computational burden of the CC-OPF optimization. The convergence of this instance is detailed in Table~\ref{fig:lambda_conv}, where values of Lagrange multipliers $\lambda_{t,b}^p$ and $\lambda_{t,b}^q$ for the TCL ensemble at at bus \# 17 are itemized for each iteration for time intervals at 8, 16, and 24 hours. Comparing the results for the 4$^{\text{th}}$ and 5$^{\text{th}}$ iterations in Table~\ref{fig:lambda_conv} reveals that desired tolerance \textcolor{black}{$\zeta$} is achieved.

\textcolor{black}{Table~\ref{table:states_comptuation} and Table~\ref{table:no_tcl_ensemble} demonstrate the computational performance of the ST-D2 algorithm for a different number of MDP states used to represent a given TCL ensemble and for a different number of TCL ensembles hosted in the distribution system, respectively. Naturally, increasing the number of states in each ensemble and the number of TCL ensembles in the system leads to greater computing times. Table~\ref{table:IEEE_system} summarizes the computing times needed to solve the proposed model using the proposed ST-D2 algorithm on the 37-, 123- and 8500-bus IEEE systems, \cite{schneider2017analytic}, with a different number of TCL ensembles. As expected, the computational time increases for a greater number of buses and TCL ensembles. However, in all cases considered in Table~\ref{table:IEEE_system}, the optimal solution is returned by the ST-D2 algorithm within an acceptable time for operational tasks. } 

\begin{table}[!t]
\centering
\scriptsize
\caption{Computational performance of the proposed ST-D2 algorithm with four TCL ensembles over 24 time intervals.}
\vspace{3pt}
\begin{tabular}{ c|cc}
\hline
\multirow{3}{*}{Iteration No.} & \multicolumn{2}{c}{Computational Time (s)} \\
\cline{2-3}
& MDP (Step 1) & CC-OPF (Step 2) \\
\hline
1 &\textcolor{black}{6.20} &36.93 \\
2 &\textcolor{black}{4.12} &42.48 \\
3 &\textcolor{black}{5.68} &43.15 \\
4 &\textcolor{black}{3.92} &47.42 \\
\hline
Total time &\qquad \qquad \qquad \textcolor{black}{189.90}\\
\hline
\end{tabular}
\label{table:cp}
\end{table}

\begin{table}[!t]

\scriptsize
\caption{Convergence of the ST-D2 algorithm for time intervals at 8, 16 and 24 hours for the TCL ensemble at bus \# 17.}
\vspace{3pt}
\begin{tabular}{ c|cc|cc|cc }
\hline
\multirow{3}{*}{Iteration No.}& \multicolumn{6}{c}{Lagrange Multipliers} \\
\cline{2-7}
& \multicolumn{2}{c|}{t=8} & \multicolumn{2}{c|}{t=16} & \multicolumn{2}{c}{t=24}\\
\cline{2-7}
& $\lambda^{p}$ & $\lambda^{q}$ & $\lambda^{p}$ & $\lambda^{q}$ & $\lambda^{p}$ & $\lambda^{q}$\\
\hline
1 & 15.4205 & 0.2575 & 15.4205 & 0.2575 & 15.4205 & 0.2575 \\
2 & 12.5518 & -0.6987 & 12.6090 & -0.6796 & 12.8543 & -0.5978 \\
3 & 12.6305 & -0.6724 & 12.6837 & -0.6547 & 12.9228 & -0.5750 \\
4 & 12.6284 & -0.6731 & 12.6818 & -0.6553 & 12.9210 & -0.5756 \\
5 & 12.6284 & -0.6731 & 12.6818 & -0.6553 & 12.9210 & -0.5756 \\
\hline 
\end{tabular}
\label{fig:lambda_conv}
\end{table}

\begin{table}[!t]
\centering
\scriptsize
\caption{\textcolor{black}{Computational performance of the proposed ST-D2 algorithm for a different number of states within a given TCL ensemble over 24 time intervals.}}
\vspace{3pt}
\begin{tabular}{ c|cc|c}
\hline
\multirow{3}{*}{\textcolor{black}{No. of States}} & \multicolumn{3}{c}{\textcolor{black}{Computational Time (s)}} \\
\cline{2-4}
& \textcolor{black}{MDP (Step 1)} & \textcolor{black}{CC-OPF (Step 2)} &\textcolor{black}{Total} \\
\hline
\textcolor{black}{4} &\textcolor{black}{5.56} &\textcolor{black}{137.08} &\textcolor{black}{142.64} \\
\textcolor{black}{8} &\textcolor{black}{19.92} &\textcolor{black}{169.98} &\textcolor{black}{189.90} \\
\textcolor{black}{12} &\textcolor{black}{35.81} &\textcolor{black}{178.68} &\textcolor{black}{214.49} \\
\textcolor{black}{24} &\textcolor{black}{122.29} &\textcolor{black}{163.41} &\textcolor{black}{285.7} \\
\hline
\end{tabular}
\label{table:states_comptuation}
\end{table}

\begin{table}[!t]
\centering
\scriptsize
\caption{\textcolor{black}{Computational performance of the proposed ST-D2 algorithm with a different number of TCL ensembles over 24 time intervals.}}
\vspace{3pt}
\begin{tabular}{ c|cc|c}
\hline
\multirow{3}{*}{\textcolor{black}{No. of TCL Ensembles}} & \multicolumn{3}{c}{\textcolor{black}{Computational Time (s)}} \\
\cline{2-4}
& \textcolor{black}{MDP (Step 1)} & \textcolor{black}{CC-OPF (Step 2)} &\textcolor{black}{Total} \\
\hline
\textcolor{black}{1} &\textcolor{black}{19.96} &\textcolor{black}{166.44} &\textcolor{black}{186.40} \\
\textcolor{black}{2} &\textcolor{black}{19.22} &\textcolor{black}{167.76} &\textcolor{black}{186.98} \\
\textcolor{black}{3} &\textcolor{black}{19.08} &\textcolor{black}{167.22} &\textcolor{black}{186.30} \\
\textcolor{black}{4} &\textcolor{black}{19.92} &\textcolor{black}{169.98} &\textcolor{black}{189.90} \\
\hline
\end{tabular}
\label{table:no_tcl_ensemble}
\end{table}

\begin{table}[!t]
\centering
\scriptsize
\caption{\textcolor{black}{Computational performance of the proposed ST-D2 algorithm for larger IEEE systems.}}
\vspace{3pt}
\begin{tabular}{ c|ccc}
\hline
\multirow{3}{*}{} & \multicolumn{3}{c}{\textcolor{black}{Total Computational Time (s)}} \\
\cline{2-4}
& \textcolor{black}{1 TCL ensemble} & \textcolor{black}{37 TCL ensemble} &\textcolor{black}{123 TCL ensemble} \\
\hline
\textcolor{black}{IEEE 37-bus} &\textcolor{black}{164.3$^*$} &\textcolor{black}{178.9} &\textcolor{black}{N/A} \\
\textcolor{black}{IEEE 123-bus} &\textcolor{black}{181.3$^*$} &\textcolor{black}{201.0$^*$} &\textcolor{black}{267.4} \\
\textcolor{black}{IEEE 8500-bus} &\textcolor{black}{611.2$^*$} &\textcolor{black}{624.7$^*$} &\textcolor{black}{902.3$^*$} \\
\hline
\end{tabular}
\begin{flushleft}
\textcolor{black}{$^*$ All TCLs are located in the electrically most remote nodes from the root node of the distribution system.}
\end{flushleft}
\label{table:IEEE_system}
\end{table}

\subsection{System Perspective}

\begin{figure}[!b]
\centering
\includegraphics[height=5cm,width=.8\columnwidth]{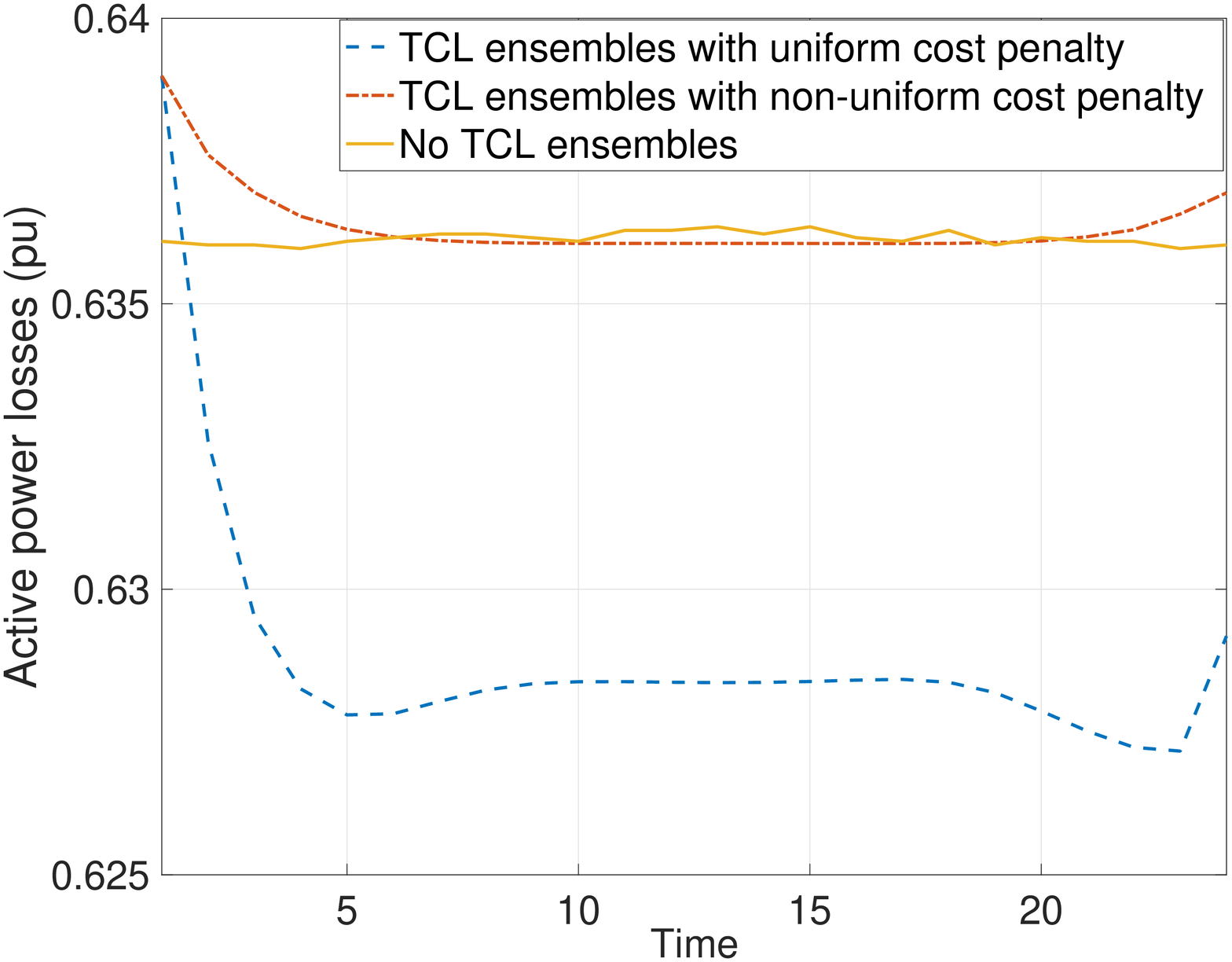}
\caption{The active power losses for the uniform and non-uniform cost penalty cases, as well as without the dispatch of TCL resources.}
\label{fig:power_losses}
\end{figure}

Solving the integrated optimization problem as in Eq.~\eqref{obj}-\eqref{int_opf} leads to the following two main results. First, it reduces the active power losses in the distribution system as explicitly formulated in the objective function. Second, it improves compliance with voltage limits. 

\begin{figure}[!t]
\centering
\includegraphics[height=5cm,width=.8\columnwidth]{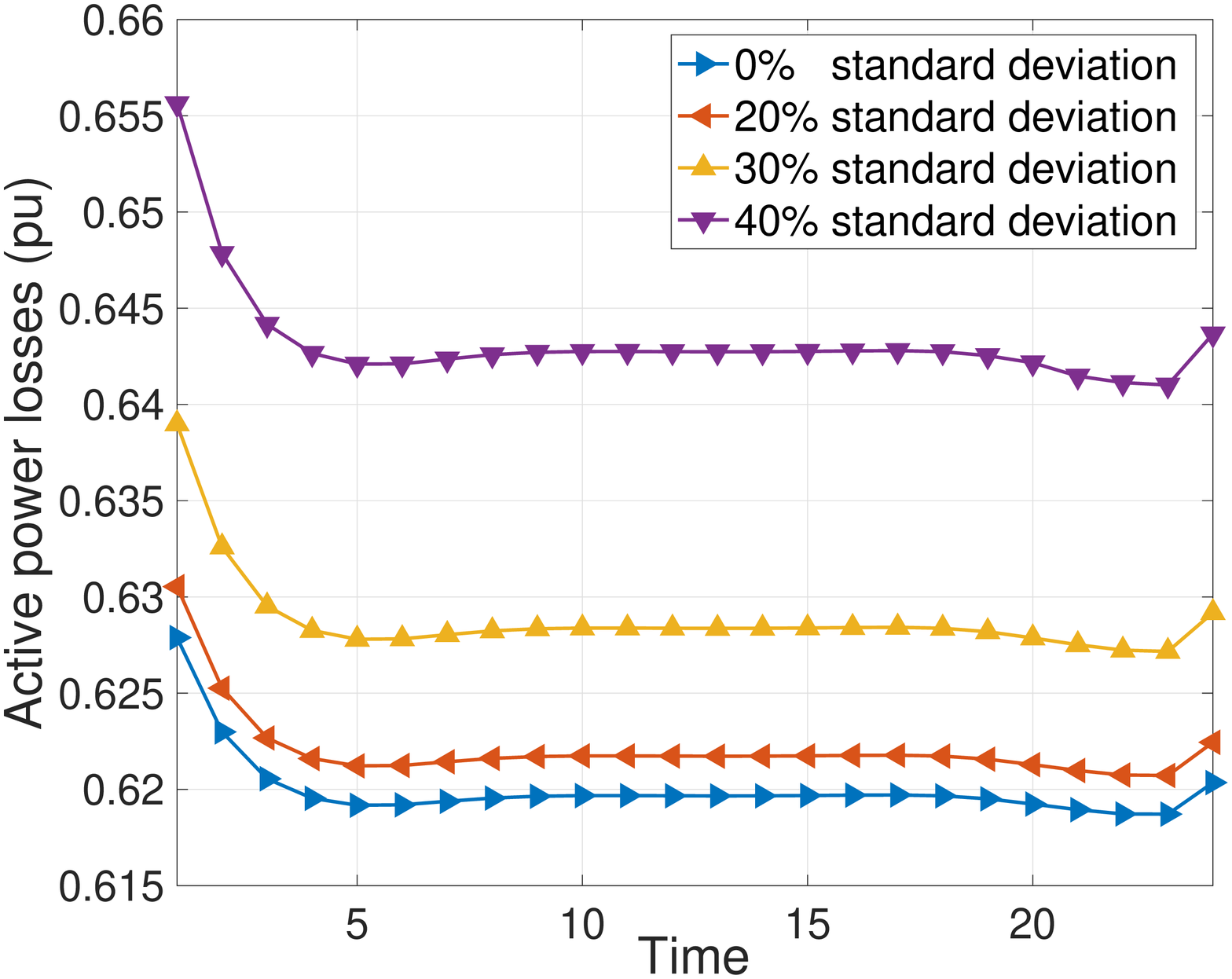}
\caption{The active power losses for different values of the standard deviation ($\sigma_{t,b}$) on the forecast error with the non-uniform cost penalty.}
\label{fig:power_losses_var}
\end{figure}

\begin{figure}[!t]
\centering
\includegraphics[height=5cm,width=.8\columnwidth]{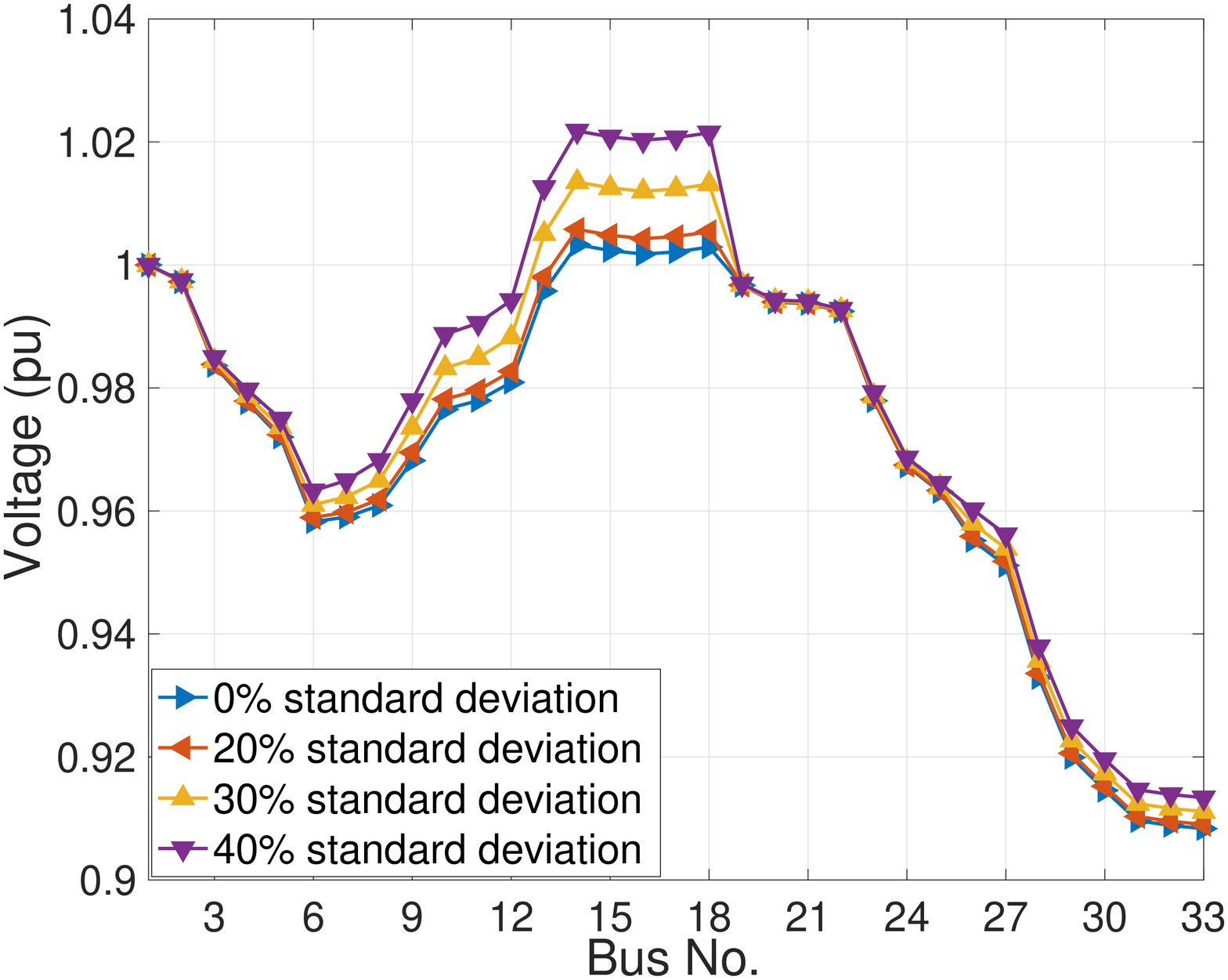}
\caption{Voltage profile for different values of the standard deviation ($\sigma_{t,b}$) on the forecast error at a given time interval ($t=24$ hours). The spikes around buses \# 14 is due to the injection of the controllable distributed generator. }
\label{fig:volt_var}
\end{figure}

  Fig.~\ref{fig:power_losses} illustrates the effect of dispatching TCL ensembles within the integrated optimization on the active power losses. If there is no MDP optimization, the losses remain constant during the optimization horizon. The uniform cost penalty that dispatches the TCLs more aggressively than the non-uniform cost penalty, as further discussed in Section~\ref{sec:case_study_tcl}, is more effective in reducing the active power losses relative to the case without the TCL resources. As shown in Fig.~\ref{fig:power_losses_var}, the effectiveness of the TCL dispatch for reducing the losses is particularly important for large standard deviations of the forecast error that cannot otherwise be dealt with efficiently using traditional controls considered in the CC-OPF formulation. Similarly, it helps improve the voltage profile as the uncertainty of the forecast error increases as shown in Fig.~\ref{fig:volt_var}.

\begin{figure}[!t]
\centering
\includegraphics[height=5cm,width=.8\columnwidth]{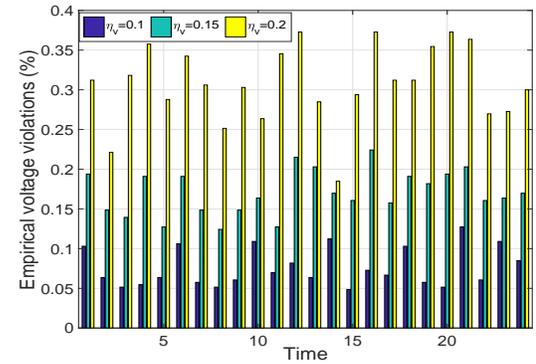}
\caption{Statistics of the voltage limit violations for different values of $\eta_{v}$ over 24 time intervals in 500 randomly generated samples. }
\label{fig:cc_vv}
\end{figure}

To further evaluate the effect of the TCL dispatch on the compliance with voltage limits enforced in Eq.~\eqref{CC_vlim1}-\eqref{CC_vlim2}, we generate 500 random samples representing the PV outputs and assess the feasibility of the solution obtained by the integrated optimization problem for different values of $\eta_v$. This assessment is performed by re-dispatching the obtained solution for each random sample. Fig.~\ref{fig:cc_vv} presents the statistics on the total number of voltage constraint violations during the optimization horizon for different values of $\eta_v$. In all instances observed in Fig.~\ref{fig:cc_vv} the empirical probability of violation is below the values postulated on $\eta_v$ in \eqref{CC_vlim1}-\eqref{CC_vlim2}. \textcolor{black}{An improvement in the compliance with voltage limits for tighter values of $\eta_v$ comes at an incremental increase in the objective function of the DSO ($<0.1\%$). }

\subsection{Perspective of TCL Ensembles} \label{sec:case_study_tcl}

\begin{figure}[!t]
\centering
\includegraphics[trim={0.5cm 0 0.5cm 0},width=\columnwidth]{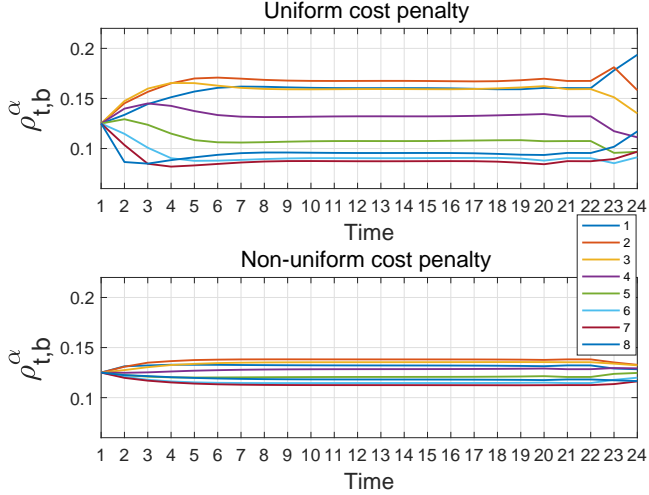}
\caption{Optimal steady-state probabilities $\rho_{t,b}^{\alpha}$ for different states $\alpha$ at the aggregator on the TCL ensemble at bus \# 26 with the uniform and non-uniform cost penalty for the transitions between different states. The legend denotes different states of the TCL ensemble as shown in Fig.~\ref{mdp:states}. }
\label{fig:prob_den}
\end{figure} 
From the perspective of the TCL ensembles, their ability to exercise dispatch flexibility mainly depends on the value of parameter $\gamma_{t,b}^{\alpha\beta}$. Fig.~\ref{fig:prob_den} compares the TCL decisions for the uniform and non-uniform cost penalty cases.
\textcolor{black}{As the penalty $\gamma_{t,b}^{\alpha\beta}$ considered in the non-uniform cost penalty setting weights out-of-cycle transition probabilities higher than next-step probabilities, the non-uniform cost penalty case will return more homogeneous transition decisions $\mathcal{P}^{\alpha\beta}_{t,b}$ of the aggregator during the optimization horizon. As a result, the steady-state probability $\rho_{t,b}^{\alpha}$ for each state will be more homogeneous.} The probability masses associated with one-step ahead transitions (largest as shown in Table \ref{Table:mdp_states}) can be exercised at a lower cost compared to the other transitions and hence lead to more homogeneity. In practice, this homogeneity implies that the \textcolor{black}{non-uniform cost penalty case} does not lead to drastic changes in the power consumption of the TCL ensemble, as per \textcolor{black}{Eq.~\eqref{mdp_injP1}-\eqref{mdp_injQ1}}, and is thus more suitable for accommodating comfort constraints of TCL users. On the other hand, the uniform cost penalty does not discriminate abrupt changes in the power consumption of the TCL ensemble and leads to more dispersed transitions. The difference between the \textcolor{black}{uniform and non-uniform cost penalty cases} presented in Fig.~\ref{fig:prob_den} indicates that there is a subtle trade-off between the comfort preferences of the TCL users driven by their default dynamics ($\overline{\mathcal{P}}^{\alpha\beta}_{t,b}$) and their ability to exercise dispatch flexibility. Both parameters $\overline{\mathcal{P}}^{\alpha\beta}_{t,b}$ and $\gamma_{t,b}^{\alpha\beta}$ can be refined within the proposed MDP optimization by using reinforcement learning \cite{7401112}. We leave it for our future work.

\section{Conclusion} \label{Sec:Conclusion}
This paper presents a modeling framework and algorithm to integrate TCL ensembles in PV-dominant distribution systems and co-optimize their dispatch flexibility with the rest of the distribution system resources. The case study demonstrates that the proposed model is capable of leveraging the dispatch flexibility of TCL ensembles to reduce active power losses and maintain nodal voltage magnitudes within an acceptable range. Comparison between the uniform and non-uniform cost penalty cases reveals that accounting for comfort preferences of TCL users can significantly influence the effect of TCL ensembles on the distribution system. The use of chance constraints on voltage limits also provides a flexible mechanism to address the conservatism of the solution and is effective in reducing violations of voltage limits.

\bibliographystyle{IEEEtran}
\bibliography{ref.bib}

\appendices
\section{Backward-Forward Algorithm} \label{sec:algoirthm}
We overview the backward-forward algorithm below: 
\begin{itemize}
\item \underline{Backward in time step.} Starting at $t=|\mathcal{T}|$, solve \eqref{MDP:obj}-\eqref{mdp_integrality} for $\mathcal{P}$ recursively backward in time, i.e. $t \rightarrow 1 $. This process returns optimal $\mathcal{P}_{t,b}^{\alpha \beta}$ for transitions to all states $\alpha$ from state $\beta$ at time $t$ given associated cost functions. Each problem can be solved either by a Lagrange relaxation or by minimizing a convex function.

\item \underline{Forward in time step.} Reconstruct $\rho$ using the relationship in \eqref{MDP_evol} forward in time, i.e. $t \rightarrow |\mathcal{T}| $, with the initial condition on $\rho_{t=0,b}^{\alpha}$ = $\rho_{in;b}^{\alpha}$,$\forall \alpha$, where $\rho_{in;b}^{\alpha}$ is given. 
\end{itemize}

Interested readers are referred to Appendix 1.9 in \cite{Chertkov_MDP} and to \cite{TCL_networks_Misha} for more details. 

\section{SOC Reformulation of the Chance Constraints} \label{Appendix:conic_refor}
Let $\xi$$\sim$$N(\mu,\Sigma)$ be the vector of random variable with the means and variances given by the vector $\mu$ and covariance-matrix $\Sigma$, respectively, and let $b$ and $x$ be the vectors of parameters and decision variables. The chance constraint of the form:
\begin{flalign}
  \mathbb{P}(\xi^T x \leq b) \geq 1 - \epsilon \label{ccgen1} 
\end{flalign}
can be represented in the following form \cite{ADMM}:
 \begin{flalign}
  \mu^Tx + \Phi^{-1}(1-\epsilon) \sqrt{x^\top \Sigma x} \leq b  \label{ccgen2} 
\end{flalign}
where $\epsilon \in [0,1/2]$ is a given tolerance to violations and $\Phi^{-1}$ is the inverse cumulative distribution function of the standard normal distribution. Eq.~\eqref{ccgen2} is then convex and equivalent to the following SOC constraint \cite{ADMM}:
\begin{align}
& t \geq \left\Vert {\Sigma^{\frac{1}{2}} x} \right\Vert_2  \label{ccgen3} \\
& \mu^Tx + \Phi^{-1}(1-\epsilon)t \leq b. \label{ccgen4}
\end{align}

\end{document}